\documentclass[aps,twocolumn,superscriptaddress]{revtex4}
\usepackage{graphicx}% Include figure files
\usepackage{dcolumn}% Align table columns on decimal point
\usepackage{bm}% bold math

\begin{document}

\title{Bose-Fermi Mixtures Near an Interspecies Feshbach Resonance:\\
 Testing a Non Equilibrium Approach}
\author{Daniele C. E. Bortolotti}
\affiliation{JILA and Department of Physics, University of Colorado, Boulder, CO 80309-0440}
\affiliation{LENS and Dipartimento di Fisica, Universit\'{a} di Firenze, and INFM, Sesto Fiorentino, Italy}
\author{Alexandr V. Avdeenkov}
\affiliation{Institute of Physics and Power Engineering, Obninsk, Russia}
\author{Christopher Ticknor}
\affiliation{JILA and Department of Physics, University of Colorado, Boulder, CO 80309-0440}
\author{John L. Bohn}
\affiliation{JILA and Department of Physics, University of Colorado, Boulder, CO 80309-0440}
 \email{bohn@murphy.colorado.edu}

\date{\today}

\begin{abstract}
We test a non equilibrium approach to study the behavior of a
Bose-Fermi mixture of alkali atoms in the presence of a Feshbach
resonance between bosons and fermions.  To this end we derive the
Hartree-Fock-Bogoliubov (HFB) equations of motion for,
the interacting system. This approach has proven very successful in the study
of resonant systems composed of Bose particles and Fermi particles.
However, when applied to a Bose-Fermi mixture, the HFB
theory fails to identify even the correct
binding energy of molecules in the appropriate limit. Through a more
rigorous analysis we are able to  ascribe
this difference to the peculiar role that bosonic depletion plays in
the Bose-Fermi pair correlation, which is the mechanism through which
molecules are formed. We therefore conclude that molecular formation
in Bose-Fermi mixtures is driven by three point and higher order
correlations in the gas, unlike any other resonant system studied in
the context of ultra-cold atomic physics.
\end{abstract}

\maketitle
\section{Introduction}
Feshbach resonances have been recently discovered in ultracold mixtures
of bosonic fermionic alkali atoms
\cite{jin,ketterle}. Together with 
the achievement of degenerate states of such systems
\cite{inguscio,ketterle2,jin2}, this experimental feat has opened investigative
opportunities for the study of new ultracold regimes.  
From the theoretical point of view, on the other hand, studies of
Bose-Fermi mixtures to date have been mostly 
limited to non 
resonant physics, focusing mainly on mean field effects in trapped
systems \cite{roth,roth2,modugno,hu,liu,adh,buc}, phases in optical
lattices \cite{albus,lewen,roth3,sanp}, or equilibrium 
studies of homogeneous gases, focusing mainly on phonon induced 
superfluidity  or beyond-mean-field effects
\cite{stoof1,heise,efre,vive,matera,gardiner}. 

This paper introduces a time dependent theory of the
Bose-Fermi  
mixture that accounts for the resonant interaction.  In systems where
the resonant 
interaction is between two bosons \cite{holland_mol,stoof2,oxford} or between
two fermions \cite{holland_ren,stoof3,victor,griffin,sasha1,oxford1},  
the theory of ``resonant superfluidity'' has already been articulated.
This theory is, so far, a big success.  In the Bose case, it quantitatively
describes the coherent conversion of bosonic atoms into bosonic molecules
and back. Indeed, Ramsey interferometry on this system, coupled with this
theoretical analysis,  has produced the
most accurate interaction potentials yet between ultracold rubidium atoms
\cite{claussen}.  In the Fermi case, the theory has produced important
qualitative insights into the crossover regime between weakly-interacting
Cooper pairs on the one hand, and Bose-condensed molecules on the other
\cite{holland_ren,stoof3,victor,griffin,sasha1,sasha2,oxford1,oxford2}. 

It seems worthwhile, therefore, to adapt the same level of theory
to the resonant Bose-Fermi mixture.  In this paper we formulate the problem
by writing down the relevant equations of motion at the level of
Hartree-Fock-Bogoliubov (HFB) approximation.
The equation of motion are suitably number- and energy-conserving, as
are their counterparts in boson of fermion systems. However, in sharp
contrast to these systems, the HFB theory applied to the Bose-Fermi
resonance does not provide quantitatively reasonable
results. Specifically we show, by direct numerical solution that the
theory cannot reproduce the binding energy of a Bose-Fermi molecule,
even in the limit of low density.

The source of this difficulty lies in the approximate treatment of
three-body correlations in the theory. The molecules, after all, are 
composed of two atoms, so the atom-atom-molecule correlation function
is of central importance in determining properties of the resulting
molecules.  In the HFB theory, this three-body correlation function 
is approximated
in terms of two-body correlation functions, which is adequate for
Bose-Bose and Fermi-Fermi resonances, but not for the Bose-Fermi
mixture.   Ultimately, the critical missing piece will turn out
to involve the non condensed bosonic atoms.

This paper is organized as follows: We begin our discussion in section
\ref{sec2}   by introducing the Hamiltonian of the system, and
justifying such choice. We then proceed to outline the
Bogoliubov-Born-Green-Kirkwood-Yvon  (BBGKY) formalism used 
to derived the HFB equations of motion, and show the form they take in
free space. In section \ref{sec3} we present our results, by first
analyzing the equations by physical and analytical insight, and then
presenting numerical results in support of our conclusions.  

Section \ref{sec4} approaches the problem from an alternative,
perturbative point of view, relevant to low fermionic densities. From
this analysis it is clear that molecular binding energies will not be
recovered without adequately accounting for the non condensed bosons,
thus pointing to their need for a higher-order theory.

\section{Theoretical Formalism}
\label{sec2}
\subsection{The Hamiltonian} 

We are interested primarily in the effects of resonant
behavior  on the otherwise reasonably understood properties of the
system. To this end we use a model which, in the last few years, has
become one of the standards  in the 
literature, and which was used to study the effect of resonant scattering in
systems composed of bosons \cite{holland_mol,stoof2,oxford}, and fermions
\cite{holland_ren,stoof3,victor,griffin,sasha1,sasha2}. Because there is already a
significant 
literature explaining the details involved in the choice of the
appropriate model Hamiltonian, we only outline the extent of the
approximation involved in such a choice.

 An accurate
approach to the problem 
would have to incorporate several scattering channels, since the resonance in
question is a consequence 
of the intertwined behavior of the complex internal structures of the
atoms. In a field theoretical sense, that would imply having to consider
vector fields for the bosons and fermions with as many components as
there are spin states involved in the interaction, and a non local
interaction tensor of adequate size, to account for all coupling
between such components.
Fortunately, if we assume that the resonances in the system are
sufficiently far from each other, such that it is possible to define a
``background,'' or away from resonance behavior, we can focus on only
one resonance at a time, which in turn makes it possible for an
effective two channel model to describe the resonance.
Furthermore, since the closed channel threshold is energetically
unaccessible at the temperatures of interest, we can ``integrate out''
the closed channel components of the fermion and boson field, in favor
of a fermion fields which we identify as representative of the motion
of one boson and one fermion, and which we dub the ``molecular field.'' In the
appropriate limit the molecular field identifies
bound states between fermions and bosons. We emphasize that
the molecular field is a theoretical artifice that
alleviates the need to treat relative motion of two atoms on the
natural scale of the interaction (tens of Bohr radii). However, this
model is appropriate for the study of the systems at hand,
typically composed of $10^{12}$ $^{40}K$ atoms per cubic
centimeter, whereby the
characteristic length scales associated with the many body system are
of the
order of 
the inverse Fermi wavenumber, (thousands of Bohr radii), and 
the average interparticle spacing (tens of thousands Bohr radii),
which is
given by $({3\over 4 \pi \rho})^{1/3}$, where $\rho$ is the atomic
density. 
Lastly, since the coupling terms in the Hamiltonian represent an
effective interaction, we can
choose its functional form, and we do so by choosing to deal with
contact interactions, which simplify the calculations immensely.

The resulting Hamiltonian has the following form:

\begin{equation}
H=H_0+H_I,
\end{equation}
where
\begin{eqnarray}
H_0&=&\sum_p \epsilon_p^F \ \hat{a}_p^{\dagger}\hat{a}_p + \sum_p
\epsilon_p^B  \
\hat{b}_p^{\dagger}\hat{b}_p  +  \sum_p \left( \epsilon_p^M + \nu
\right) \
\hat{c}_p^{\dagger}\hat{c}_p \nonumber \\
&+&{\gamma \over 2 V} \sum_{p,p',q}
\hat{b}_{p-q}^{\dagger}\hat{b}_{p'+q}^{\dagger} \hat{b}_{p} \hat{b}_{p'}
\nonumber \\ 
H_I &=& {V_{bg} \over V} \sum_{p,p',q}
\hat{a}_{p-q}^{\dagger}\hat{b}_{p'+q}^{\dagger} \hat{a}_{p} \hat{b}_{p'}
\nonumber \\
&+&{g \over
  \sqrt{V}}\sum_{q,p}(\hat{c}_q^{\dagger}\hat{a}_{-p+q/2}\hat{b}_{p+q/2} +
h.c.). \nonumber \\
\label{act-bfm}
\end{eqnarray}
Here $\hat{a}_p,\hat{b}_p$, are the
annihilator operators for, respectively, fermions and bosons, $\hat{c}_p$
is the annihilator operator for the molecular field
\cite{holland_ren,sasha1,sasha2}; $\gamma = 4\pi a_b / m_b$ is the interaction
term for  
bosons, where  $a_b$ is the boson-boson scattering
length; and $V_{bg},\nu$, and $g$ are
parameters related to the Bose Fermi interaction, yet to be  
determined. Also we define single particle energies
$\epsilon^{\alpha} = p^2 / 2 m_{\alpha}$, where 
  $m_{\alpha}$ indicates the mass of bosons, fermions, or pairs, 
and $V$ as the  volume  of a quantization box with periodic boundary
conditions. 
\subsection{Two Body Scattering Parameters}

The first step is to find the values for $V_{bg},\nu,g$, in terms of
measurable parameters. We will, for this purpose, calculate the 2-body
T-matrix resulting from the Hamiltonian in
eq. \ref{act-bfm}. Integrating the molecular field out of the real
time path integral, \cite{negele} leads to the following Bose-Fermi
interaction Hamiltonian 
\begin{equation}
H_I^{2body}={1 \over  V} \left( V_{bg} + {g^2 \over E - \nu}\right)
\sum_{p} \hat{a}_{p}^{\dagger}\hat{b}_{-p}^{\dagger} \hat{a}_{p}
\hat{b}_{-p} .
\end{equation}
This expression is represented in center of mass coordinates, and $E$ is the
collision energy of the system. From the above equation we read
trivially the zero energy T-matrix in the saddle point approximation
\begin{equation}
T = (V_{bg} -{g^2\over\nu}),
\label{t2b}
\end{equation}
which corresponds to the Born approximation, and we proceed to match
it to the conventional parameterization 
\cite{holland_ren,stoof2} 
\begin{equation}
T(B) = {2 \pi \over m_{bf}} a_{bg} \left(1-{\Delta_B
\over (B - B_0)} \right),
\label{e0param}
\end{equation} 
where $a_{bg}$ is the value of the scattering length
far from resonance, $\Delta_B$ is the width, in magnetic
field, of the resonance, $m_{bf}$ is the reduced mass, and  $B_0$ is
the field at which the resonance is centered.

The identification of parameters between eqns, (\ref{t2b}) and
(\ref{e0param}) proceeds as follows: far from resonance, $|B-B_0|
>>\Delta_B$, the interaction is defined by a background scattering
length, via $V_{bg}={2 \pi a_{bg} \over m_{bf}}$. To relate magnetic
field dependent quantity $B-B_0$ to its 
energy dependent analog $\nu$, requires defining a parameter
$\delta_B=\partial \nu / \partial B$, which may be thought of as a
kind of magnetic moment for the molecules.
It is worth noting that $\nu$ does not
represent the position of the resonance nor the binding energy of the
molecules, and that, in general $\delta_B$ is a field-dependent
quantity, since the molecular binding energy approaches threshold
quadratically. For current purposes we identify $\delta_B$ by its
behavior far from resonance, where it is approximately constant. Careful
calculations of scattering properties using the model in
eq. (\ref{act-bfm}), however,
leads to the correct Breit-Wigner behavior of the 2-body T-matrix, as
we show in section \ref{scatsec}

Finally we get the following identifications:
\begin{eqnarray}
&V_{bg}={2 \pi a_{bg} \over m_{bf}} \nonumber \\
&g=\sqrt{V_{bg} \delta_B \Delta_B} \nonumber \\
&\nu=\delta_B (B - B_0).\nonumber \\
\end{eqnarray}
For our calculations we use the 511G resonance in the
$^{40}$K-$^{87}$Rb system, 
the parameters we use in the calculations to follow are
$a_{bg}=-202 a_0$,  $\delta_B=5.1 \times 10^{-5}$ K/G, and $\Delta_B=1$G.

\subsection{The Formalism}

We now move on to the many body analysis, and derive the Heisenberg
equations of motion for the many body system. The way this is done, is
to find equation of motion for correlation functions, $f_s(x_1,....x_s)$, which
represent the probability of finding $s$ particles at positions $x_1,
..., x_s$. As it turns out, the equation of motion
for the correlation function $f_1$ will depend on the function $f_2$,
which in turn will depend on $f_3$, and so on all the way to $f_N$,
where $N$ is the total number of particles in the system. This is known
as a Bogoliubov-Born-Green-Kirkwood-Yvon (BBGKY) hierarchy \cite{huang}.
In practice we will be concerned with momentum space correlation
functions, but the idea is the same. 

Given the large number of particles in the system, it is 
impossible to calculate equation of motions for all correlation
functions, and we need to invoke an approximation.
In practice, correlation functions are often calculated
only up to two-body correlations, $s=2$.  This is justified under
the assumption that interactions are suitably ``weak.''  Higher-order
correlations are included in an approximate way by considering, not
the actual atomic constituents, but rather combinations
called quasiparticles.  The quasiparticles are defined to be noninteracting,
so that their higher-order correlation functions can be written
in terms of second order correlation functions \cite{negele}.

Using this qualitative idea we proceed to develop a more formal
understanding. In statistical field theory, given an operator
$\mathcal{O}$, and Hamiltonian $H$, we define the thermal average
of $\mathcal{O}$ with respect to $H$ as
$<\mathcal{O}>_H=1 / Z Tr\left\{\mathcal{O} e^{-\beta H}\right\} $,
where $\beta=(k_B T)^{-1}$ 
is the inverse temperature, and $Z=Tr\left\{e^{-\beta H}\right\}$ is
the partition 
function. In this framework, the 1-particle correlation 
function is defined as the thermal average of the number operator,
with respect to the Hamiltonian of the system. 

In the quasi-particle representation, we define the annihilation
operator for quasi-particles as $\alpha$, reminding ourselves that it
is a complicated function of
$a,a^{\dagger},b,b^{\dagger},c,c^{\dagger}$. In momentum space, the 1-particle 
correlation function  in this representation, will then be
$<\alpha_{p1} \alpha_{p2}>_{H_{qp}}$, where $H_{qp}$ is the
(noninteracting) quasi-particle Hamiltonian.

Now we introduce the real approximation, namely that the quasi-particles
can be written as linear combinations of all possible products of two
operators (except for averages involving one fermionic and one bosonic
operator, which are easily shown to vanish). 
The procedure is then to find the Heisenberg equations of motion for
these pairs 
of operators, and then averaging over the quasi-particle Hamiltonian
\begin{equation}
i\hbar {\partial \over \partial t} <\mathcal{O}>_{H_{qp}} = <[\mathcal{O},H
]>_{H_{qp}},
\end{equation}
which being Gaussian allows us to invoke Wick's theorem to decompose
all higher order correlations in 1-particle correlations, thus
truncating the BBGKY hierarchy. 

\subsection{The Equations of Motion}
Before generating Heisenberg equations, we need to
take a little care in the treatment of the Bose field, to properly
treat the condensed part. To this end we perform the usual separation
of mean-field and fluctuations of the Bose field, substituting $b_0$ (
the zero-momentum component of the Bose gas) with a c-number
$\phi=<b_0>_{H_{qp}}$ ,
and identifying it with the condensate amplitude, while $<b_{p \neq
  0}>_{H_{qp}} =0$ are the fluctuations. We insert these definitions in
  the Hamiltonian in eq. (\ref{act-bfm}), then proceed to calculate
  commutators. 

Since we wish to limit our analysis to a homogeneous gas, we note that
the correlation functions $f_1(x,x')$ can be written in terms of a
relative coordinate $y=x-x'$. Thus in momentum space $f_1(p)$ is the
probability to find a particle with momentum p in the gas, or in other
words it is the momentum distribution of the system. 

Having taken all appropriate commutators, and applied Wick's theorem,
(for more details on the procedure see \cite{burnett}, or
Appendix \ref{Appendix1} for the derivation of a sample equation.), we
obtain the 
following self consistent set of equations of motion for the system:
\newcounter{eq3}
\stepcounter{equation}
\setcounter{eq3}{\theequation}
\setcounter{equation}{0}
\renewcommand{\theequation}{\arabic{eq3}.\alph{equation}}
\begin{widetext}
\begin{eqnarray}
i\hbar {\partial \over \partial t} \phi &=& V_{bg} \rho_F \phi +
  \gamma\ (2 \phi \tilde{\rho}_B + \Delta_B \phi^*) + g \rho_{MF}^* +
  \gamma  |\phi|^2 \phi \label{eqm1} \\
\hbar {\partial \over \partial t} \tilde{\eta}(p)&=& 2 \gamma \ \Im m
\left[ \kappa_B(p) (\phi^{*^2}+\Delta_B^*)\right] \label{eqm2} \\
i\hbar {\partial \over \partial t} \kappa_B(p)&=&\left[\epsilon_P^B +
  2 V_{bg} \rho_F+ 4 \gamma
  (|\phi|^2+\tilde{\rho_B})\right]\kappa_B(p)+\gamma
 (2\tilde{\eta_B}(p)+1)(\phi^2+\Delta_B)\label{eqm3}\\ 
\hbar {\partial \over \partial t} \eta_F(p)&=&-2 g \ \Im m  (\phi
  \eta_{MF}(p))\label{eqm4}\\
i\hbar {\partial \over \partial t} \kappa_F(p)&=&\left[\epsilon_p^F +
  2 V_{bg} (\tilde{\rho_B}+ |\phi|^2)\right]\kappa_F(p)\label{eqm5}\\
\hbar {\partial \over \partial t} \eta_M(p)&=& 2 g \ \Im m  (\phi
  \eta_{MF}(p))\label{eqm6}\\
i\hbar {\partial \over \partial t}
  \kappa_M(p)&=&\left[\epsilon_P^M+\nu\right] \kappa_M(p)
  \label{eqm7}\\
i \hbar {\partial \over \partial t} \eta_{MF}(p)&=&\left[\epsilon_p^F-
  \epsilon_p^M -\nu+ V_{bg} (\tilde{\rho_B}+ |\phi|^2)\right]
  \eta_{MF}(p) -g \phi^*\left(\eta_F(p)-\eta_M(p)\right) \label{eqm8}\\
i \hbar {\partial \over \partial t} \kappa_{MF}(p)&=&\left[\epsilon_p^F+
  \epsilon_p^M +\nu+ V_{bg} (\tilde{\rho_B}+
  |\phi|^2)\right]\kappa_{MF}(p)-g\left[\phi 
  \kappa_F (p)+\phi^*\kappa_M (p)\right],
\label{eqm9}
\end{eqnarray}
\end{widetext}
\renewcommand{\theequation}{\arabic{equation}}
\setcounter{equation}{\arabic{eq3}}
where $\tilde{\eta_B}(p)=<b_{p\neq 0}^{\dagger} b_{p\neq 0}>_{H_{qp}}$
is the momentum 
distribution of non-condensed bosons, and $\tilde{\rho}_B=\int {dp \over 2
  \pi^2} p^2  \tilde{\eta_B}(p)$ is the density of non-condensed
bosons; $\kappa_B (p)=<b_{p\neq 0} b_{p\neq 0}>_{H_{qp}}$ is the
anomalous distribution of bosonic fluctuations, and $\Delta_B= \int
{dp \over 2   \pi^2} p^2  \kappa_B(p)$ the anomalous density. Similarly
$\eta_{F,M}(p)$ are 
  the fermionic and molecular distributions, $\rho_{M,F}$ the
  densities, and $\kappa_{F,M}(p)$, and $\Delta_{F,M}$ the anomalous
  molecular and fermionic distributions and densities. Finally
 $ \eta_{MF}(p)=<c_{p}^{\dagger} a_{p}>_{H_{qp}}$ and  
$  \kappa_{MF}(p)=<c_{p} a_{p}>_{H_{qp}}$    are the normal and
  anomalous distribution for molecule-fermion correlation, with the
  associated densities $\rho_{MF}$ and $\Delta_{MF}$.

\section{Analysis and Results}
\label{sec3}
Equations (\ref{eqm1}-\ref{eqm9}) describe the complete self-consistent set of HFB
equations for the resonant BF mixture.
Inspection of these equations, however, allows us to simplify
the set quite dramatically, without sacrificing almost any of the
physics thereby contained. First, we notice that the evolution
of the anomalous fermionic densities $\kappa_{MF}(p)$,
$\kappa_{F}(p)$, and $\kappa_{M}(p)$ is entirely decoupled from the
evolution of all other quantities, and can therefore be considered
separately.
 This implies that, since we are mainly interested
in the evolution of the normal densities, we can eliminate without
approximation all the anomalous ones. 

The next thing we notice is that the evolution of the normal and
anomalous bosonic averages is completely independent of the resonant
interaction, and is controlled only by the background interactions
between bosons and with fermions. For typical background interaction
strengths, and cold enough temperatures, it is well established that
quantum depletion is minor, and 
the system is well described at the Gross-Pitaevskii level of
approximation.

We can therefore write the following reduced set of equations:
\stepcounter{equation}
\setcounter{eq3}{\theequation}
\setcounter{equation}{0}
\renewcommand{\theequation}{\arabic{eq3}.\alph{equation}}
\begin{widetext}
\begin{eqnarray}
i\hbar {\partial \over \partial t} \phi &=& (V_{bg} \rho_F+\gamma
  |\phi|^2) \phi + g \rho_{MF}^* \label{eqr1} \\
\hbar {\partial \over \partial t} \eta_F(p)&=&-2 g \ \Im m  (\phi
  \eta_{MF}(p))\label{eqr2}\\
\hbar {\partial \over \partial t} \eta_M(p)&=& 2 g \ \Im m  (\phi
  \eta_{MF}(p))\label{eqr3}\\
i \hbar {\partial \over \partial t} \eta_{MF}(p)&=&\left[\epsilon_p^F-
  \epsilon_p^M -\nu+ V_{bg} |\phi|^2\right]
  \eta_{MF}(p) -g \phi^*\left(\eta_F(p)-\eta_M(p)\right) \label{eqr4}.
\end{eqnarray}
\end{widetext}
\renewcommand{\theequation}{\arabic{equation}}
\setcounter{equation}{\arabic{eq3}}
Together with the prospect of simulating time dependent experiments, 
such a set of equations allow us to calculate many characteristics of
the system, which we could use to understand further physics or, more
importantly at this stage,  to test the theory against our knowledge
of the system in various limits. 

A relevant quantity we can calculate to this end is the binding energy of the
molecules. This can be done by an instantaneous jump of the detuning
from large and positive values, where we know the equilibrium distributions
very well, to some other arbitrary value. The system thus perturbed
oscillates at a specific characteristic frequency, which identifies
as the (unique) 
pole of the HFB many body T-matrix of the system. For negative
detunings, as shown below, this pole corresponds to the binding
energy of the molecules, dressed by the interactions in the system. 

\begin{figure}[ht]
\rotatebox{270}{\resizebox{3.5in}{!}{\includegraphics{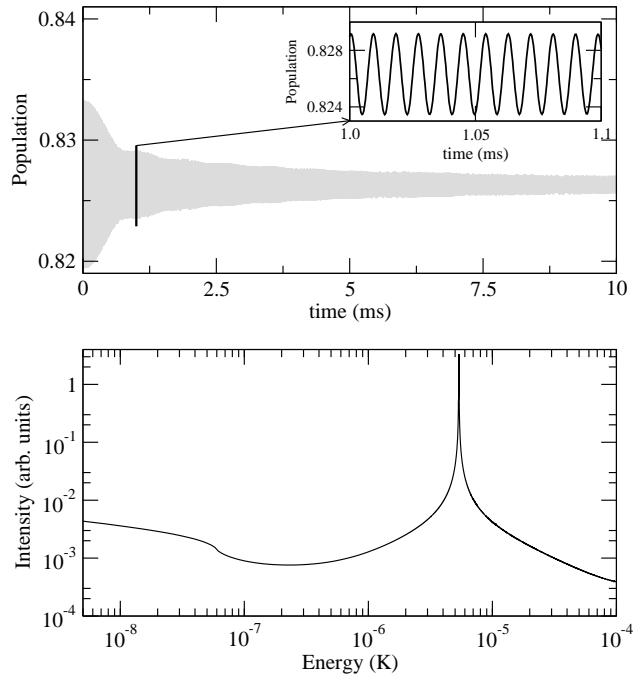}}}
\caption{The top panel represents the time evolution of the population
  of condensed atoms after detuning is suddenly shifted from infinitely
  positive to -5.1 K ($-.1 G$ magnetic detuning). The bottom panel
  shows the absolute value of 
  Fourier transform of said time evolution. the  main peak in this graph
  represents the computed value of the binding energy, which we see is
  about $5.4 10^{-6} K$. The system under consideration is composed of
  fermionic densities of  $10^{12}  cm^{-3}$, for a constant density
  ratio of five bosons per fermion.
\label{ex}}
\end{figure}

Figure \ref{ex} shows a representative example of  time evolution of
the condensate population 
(number conservation guarantees that all three populations oscillate
with the same frequency) under the conditions described above. In this
particular example, at time
t=0 the detuning is suddenly shifted to $-5.1 10^{-6} K$,
  corresponding to a magnetic field detuning of approximatively $.1
  G$. 
The response of the population shows an envelope function, indicated
by the gray shaped area, that arises from nonlinearities in the
equations of motion. The inset shows that under this envelope is a
well defined sinusoidal oscillation.

The nearly monochromatic character of the response is made clear by
Fourier transforming the time dependent population. The Fourier
Transform shown in the second panel of Fig \ref{ex} is strongly
peaked at $5.4 \times 10^{-6} K$.
Similarly, the position of the peak in the frequency
spectrum, for different final detunings, should map 
the molecular binding energy as a function of magnetic
field.

%\begin{widetext}
%\begin{center}
%\end{center}
%\end{widetext}

\begin{figure}[ht]
\resizebox{3.5in}{!}{\includegraphics{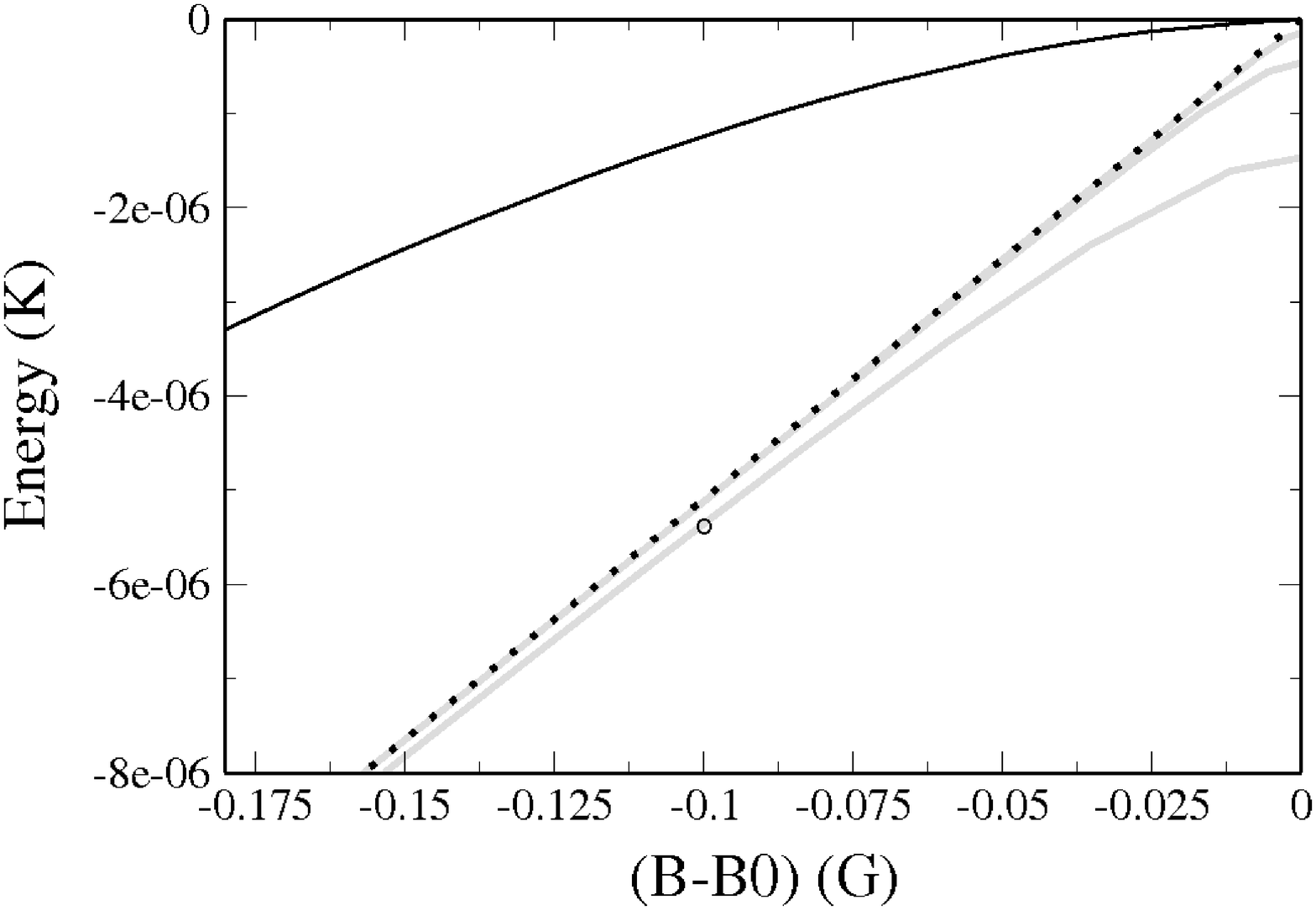}}
\caption{Plot representing the poles of the scattering t-matrix for
  the 511G  $^{87}Rb$ - $^{40}K$ Feshbach resonance. The dotted line
  represents the ``bare'' molecular detuning as a function of field, as
  defined in the text. The solid black line is the correct binding
  energy of the molecular state, obtained by means of full close coupling
  calculations, while the gray solid lines are the eigenenergies
  obtained from equations \ref{eqr1}-\ref{eqr4}, for different atomic
  densities. From top to bottom on the right the grey lines refer to
  fermionic densities of $10^{10} cm^{-3}$,$10^{11} cm^{-3}$, and $10^{12}
  cm^{-3}$, for a constant density ratio of five bosons per
  fermion. We note that for lower and lower densities the calculated
  binding energy incorrectly approaches the bare detuning instead of
  the correct two-body binding energy.
\label{freq}}
\end{figure}

Figure \ref{freq} shows the results obtained by this method. 
This plot represents the binding
energy of the molecules, dressed by the interactions in the
system. This dressing is expected to be weaker  for smaller 
densities of atoms and molecules. In
this limit, we should thus recover the two body molecular binding
energy, which can be calculated quite
accurately from two body close 
coupling calculations (solid line in fig \ref{freq}). Instead we see
that the pole behavior 
approaches the bare detuning (dashed line in fig \ref{freq}),
indicating that the renormalization of 
the binding energy obtained at the presented level of approximation is
inadequate to correctly include the two-body physics. This behavior
is in sharp contrast to the Bose-Bose resonant interaction, where the
correct binding energy is preserved at the HFB level
\cite{holland_mol}. This is also true for the Fermi-Fermi case
\cite{holland_priv}.

This discrepancy is due to the fact that the creation of molecules
requires the formation of correlations between bosons and fermions,
which, as shown in the following, cannot exist
if the density matrix is assumed to be Gaussian. Specifically what is
required is a more careful consideration of the noncondensed bosons

\section{The Importance of Bosonic Depletion}
\label{scatsec}
The reason for the failure of the HFB theory is not immediately clear
from the theory itself. To bring out the inadequacy of this theory in
the dilute limit, we now recast the problem in an alternative
perturbative form that can reproduce the correct behavior in the two
body limit. This path integral approach will also lay bare the role
of noncondensed bosons.

What we will see in the upcoming analysis may be qualitatively
understood in the following simple terms. A molecule in the gas can
decay into a pair of ``virtual'' (i.e. non energy conserving) atoms,
which can then meet again and reform the molecule. The incidence of
these events modifies the behavior of the molecule, and an
appropriate treatment of these virtual excitations is therefore
necessary to correctly include the two body properties of the
molecules in the many body theory. In
particular, the molecules can decay forming a virtual non condensed
boson, and the contribution of this set of events to the physics of the
molecules turns out to be very important. An appropriate theory would
therefore consider the coupling of the molecules to non condensed
bosons explicitly, which implies that one has to include in the
equations of motion three point averages, such as
$<c^{\dagger}_q a_{-p+q/2} b_{(p+q/2)}>$. Since the HFB theory
disregards three point averages, it only contains molecule-atom-atom
couplings of the form $<c^{\dagger}_q a_{q}> \phi_0$, where
molecules can only decay forming a condensed boson.

It is straightforward to see that the HFB theory treats 
3-body correlation functions differently depending on the 
quantum statistics of the constituents.  For a Bose-Bose
mixture, the correlation function is approximated (schematically) by

\begin{equation}
\langle bbm \rangle \approx \langle m \rangle \langle b(-q) b(q) \rangle
+2 \langle b \rangle \langle bm \rangle.
\end{equation}
The first term of the right of this expression allows explicitly for
virtual bosonic pairs of arbitrary momentum, provided that the
molecular field $\langle m \rangle$ accounts for most of the
molecules, which is assumed to be the case.  Similarly, in a mixture 
of distinct fermions, the correlation function reads
\begin{equation}
\langle f_1 f_2 m \rangle \approx \langle m \rangle 
\langle f_1(q) f_2(-q) \rangle,
\end{equation}
and the same argument applies, since the molecules are bosons.

For the Bose-Fermi mixture, on the other hand, the correlation
function would be approximated by
\begin{equation}
\langle bfm \rangle \approx \langle b \rangle\langle fm \rangle
+ \langle f \rangle \langle bm \rangle
+ \langle m \rangle \langle bf \rangle.
\end{equation}
The required virtual atom-atom pairs would arise from the third
term on the right-hand-side of this expression.  However, these
molecules are fermions, which have no mean field, $\langle m \rangle=0$.
The only surviving term is then the first one, which accounts
only for condensed bosons, and somehow correlates the fermionic
atoms to the fermionic molecules.  This is only an indirect way
to get the bosons and fermions correlated.

\subsubsection{Two-Body Scattering}
The perturbative analysis begins by recasting the Hamiltonian 
in eq. (\ref{act-bfm}) in terms of 
a 2-body action, in center of mass coordinates :
\begin{widetext}
\begin{equation}
S[\psi,\psi^{\dagger},\phi,\phi^{\dagger},\xi,\xi^{\dagger}]=S_B[\phi,\phi^{\dagger}]+S_F[\psi,\psi^{\dagger}]+S_M[\xi,\xi^{\dagger}]+S_C[\psi,\psi^{\dagger},\phi,\phi^{\dagger},\xi,\xi^{\dagger}],
\label{act-bfm2}
\end{equation}
where the field $\phi$ represents the bosons, $\psi$ the fermions, and
$\xi$ the 
fermionic molecules, and where

\begin{eqnarray}
 &S_B[\phi,\phi^{\dagger}]&=\int {d \omega \over 2 \pi}  \sum_{\bf
  p}(-\hbar \omega+
  \epsilon_p^B ) \ 
\phi^{\dagger}_{\omega,{\bf p}}  \phi_{\omega,{\bf p}} + {1 \over 
  2 V} \gamma \int {d \omega \over 2 \pi }
\sum_{{\bf p} ,{\bf p'}{\bf q}}  \phi_{\omega,{\bf p}
  -{\bf q}}^{\dagger}\phi_{\omega,{\bf p'}+{\bf q}}^{\dagger} \phi_{\omega,{\bf
 p'}} \phi_{\omega,{\bf p}} \nonumber \\ 
 &S_F[\psi,\psi^{\dagger}]&= \int {d \omega \over 2 \pi}  \sum_{\bf
  p}(-\hbar \omega+
  \epsilon_p^F ) \ 
\psi^{\dagger}_{\omega,{\bf p}}  \psi_{\omega,{\bf p}}\nonumber \\
 &S_M[\xi,\xi^{\dagger}]&= \int {d \omega \over 2 \pi}  \sum_{\bf
  p}(-\hbar \omega+
  \epsilon_p^M +\nu) \ 
\xi^{\dagger}_{\omega,{\bf p}}  \xi_{\omega,{\bf p}} \nonumber \\
 &S_C[\psi,\psi^{\dagger},\phi,\phi^{\dagger},\xi,\xi^{\dagger}]&=
 \ {V_{bg}\over V}\int {d \omega \over 2 \pi} \sum_{{\bf p} ,{\bf p'}{\bf q}}  \psi_{\omega,{\bf p}
  -{\bf q}}^{\dagger}\phi_{\omega,{\bf p'}+{\bf q}}^{\dagger} \psi_{\omega,{\bf
 p'}} \phi_{\omega,{\bf p}}
\ + g \int {d \omega \over 2 \pi} \sum_{{\bf p} {\bf
  q}}(\xi^{\dagger}_{\omega,{\bf p}}
\psi_{\omega, {\bf q}-{\bf p}} \phi_{\omega, {\bf p}} + c.c),
\label{act-desc}
\end{eqnarray}
\end{widetext}
where $\hbar \omega$ is the frequency associated with the motion of
the various fields.

\begin{figure}[ht]
\resizebox{3.5in}{!}{\includegraphics{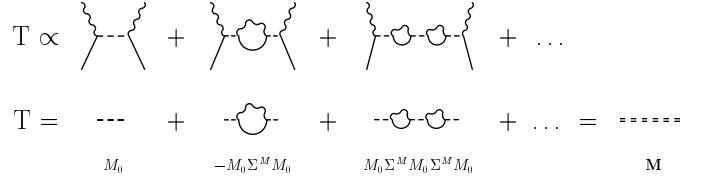}} 
\caption{Feynman diagrams representing the resonant collision of a
  fermion and a boson. Straight lines represent fermions, squiggles
  bosons, and dashed lines represent the effective composite fermions
\label{2-bdiag1}}
\end{figure}

As before we will then proceed to integrate out the molecular degree
of freedom \cite{negele} to get :
\begin{widetext}
\begin{eqnarray}
&S'_C[\psi,\psi^{\dagger},\phi,\phi^{\dagger}]&=
\int {d\omega \over 2\pi}{d\omega' \over 2\pi} \sum_{\bf p p'}\left(V_{bg}+{g^2 \over E-\nu}\right)
\phi^{\dagger}_{\omega,  \bf p}\psi^{\dagger}_{E-\omega,- \bf p}
\phi_{\omega', \bf p'}\psi_{E-\omega',- \bf p'} 
\label{vbare}
\end{eqnarray}
where E is the collision energy between the fermions and the
bosons. We then
undergo the inverse transformation to obtain:

\begin{eqnarray}
&S''_M[\xi,\xi^{\dagger}]&=-\int {d\omega \over 2\pi} \sum_{\bf p}
  \left(V_{bg}+{g^2 \over w-{p^2 \over 2 (m_b
    + m_f)} -\nu} \right)^{-1}\xi'^{\dagger}_{\omega, \bf p}\xi'_{\omega,
   \bf p} \nonumber \\
&S''_C[\psi,\psi^{\dagger},\phi,\phi^{\dagger},\xi,\xi^{\dagger}]&=
\int {d\omega \over 2\pi}\sum_{\bf p}(\xi'^{\dagger}_{E, \bf 0}
\psi_{E-\omega,-{\bf p}}\phi_{\omega,{\bf p}} + c.c).
\nonumber \\ 
\label{spp}
\end{eqnarray}
\end{widetext}
Here $\xi'$ represents the new effective (i.e. primed) molecules.
The first line of fig. \ref{2-bdiag1}
shows the diagrams describing the resonant collisions between bosons
and fermions. Here the continuous lines refer to fermions, the
squiggles to bosons, and the broken lines to effective molecules. Since we are
looking for poles of the S-matrix, we can disregard the trivial
fermion and boson propagators, and proceed, as outlined in fig.
\ref{2-bdiag1}, to calculate the renormalized propagator for $\xi'$,
denoted as 
${\bf M}$, represented there as a heavy broken line. This
object coincides with the T-matrix of the system, and shares its
poles. Using the definition of the retarded molecular self energy
$\Sigma^M$ given in fig. \ref{2-bdiag1}, and calling the molecular 
propagator $M_0$ (again for $\xi'$), we get the following Dyson series:
\begin{eqnarray}
&&T = {\bf M}=M_0-M_0\Sigma^M M_0-M_0\Sigma^M M_0\Sigma^M M_0 +\dots
  \nonumber \\
=&&M_0-M_0\Sigma^M{\bf M},
\end{eqnarray}
where $T$ is the T-matrix for the collision, and which has formal solution
\begin{equation}
T = {\bf M}={1\over M_0^{-1} + \Sigma^M}.
\label{Dyson}
\end{equation}
%\begin{widetext}
%\begin{center}
%\end{center}
%\end{widetext}

\begin{widetext}
These quantities take the explicit form 
\begin{eqnarray}
& M_0(E) &=\left(V_{bg}+{g^2 \over E -\nu} \right) \nonumber \\
& \Sigma^M(E)&=i \int \ {d\omega \over 2 \pi}{d{\bf p} \over (2 \pi)^3}{1
    \over (\hbar \omega - {p^2 \over  2 m_b} + i0^+)(E- \hbar\omega -
	  {p^2 \over 
      2 m_f} + i0^+)}   \nonumber \\ 
& & \approx -i{ m_{bf}^{3/2}\over \sqrt{2} \pi} \sqrt{E}-{ m_{bf}
    \Lambda \over \pi^2},
\label{M-E}
\end{eqnarray}
\end{widetext}
where  $m_{bf}$
is the boson-fermion reduced mass, and $\Lambda$ is an ultraviolet
momentum cutoff needed to hide the unphysical nature  of the contact
interactions; we will dwell more on that shortly. Finally
inserting eq. \ref{M-E} into eq. \ref{Dyson}, we obtain the following expression
for the T-matrix:
\begin{equation}
T(E)=\left[ {1 \over V_{bg} +{g^2 \over E - \nu}}+i{ m_{bf}^{3/2}\over
    \sqrt{2} \pi} \sqrt{E}+{ m_{bf} \Lambda \over \pi^2} \right]^{-1}
\label{tmat}
\end{equation}

To show that this expression correctly represents the two body
T-matrix for two channel resonant scattering, we will compare it to the
results we know from standard theoretical treatments, \cite{landau},
which  teach us that: 
\begin{equation}
T(E)={\pi i \over m_{bf} \sqrt{2 m_{bf} E}}\left(S(E)-1\right),
\label{Fesh}
\end{equation}
where $S(E)$ is the S matrix given by
\begin{equation}
S(E)=e^{-2 i \sqrt{2 m_{bf} E} \ a_{bf}}\left(1-{2 i \Gamma \sqrt{E},
  \over E- \epsilon_0 + i \Gamma \sqrt{E} }\right).
\end{equation}
Here $\Gamma \sqrt{E}$ is the width of the resonance, $\epsilon_0$ is a
shift associated with the detuning with respect 
to threshold of the resonance, and $a_{bf}$ is the s-wave scattering
length for the boson-fermion collision; all of these quantities can be
extracted from experimental observables, through accurate two-body
scattering calculations.

From the parametrization of the zero energy T-matrix in
eq.(\ref{e0param}), and the $E \rightarrow 0$ limit of (\ref{Fesh}), we
easily derive $\Gamma=\sqrt{2 m_{BF}} a_{BF} \delta_{\mu} \Delta
B$. With these definitions we can relate equations (\ref{Fesh})
and (\ref{tmat}), to find a regularization scheme for the theory, by
substituting the non observable parameters $g$, $\nu$, and $V_{bg}$ by
the $\Lambda$ 
dependent (renormalized) quantities $\bar{g}$, $\bar{\nu}$, and
$\bar{V_{bg}}$, such that the observable T-matrix will not be itself $\Lambda$
dependent     

Following \cite{holland_ren} we compare equations (\ref{tmat}) and  
(\ref{Fesh}), in the limit $E \rightarrow 0$, where we have (once we
include the definitions of the bare quantities)
\begin{equation}
V_{bg} - {(g)^2 \over \nu^0}=\left[ {1 \over \bar{V_{bg}} -{\bar{g}^2 \over
	\bar{\nu}}}+{ m_{bf} \Lambda \over \pi^2} \right]^{-1}.
\label{ren-cond}
\end{equation}
Since we have one equation and three unknowns, we will have to insert
some physics in the system, analyzing it one limit at a time. The
first limit is far from resonance, where $\nu\rightarrow\infty$
\begin{equation}
\bar{V_{bg}}=V_{bg} \left(1 \over 1- {m_{bf} \Lambda V_{bg} \over
  \pi^2} \right) 
\label{vbg_renorm}
\end{equation}

We are now left with the task of defining the resonant quantities, and
we have no more leeway to make physically motivated
simplifications. The equations which remain are ambiguous, which leaves
us with a set of possibilities for the choice of $\bar{g}$ and
$\bar{\nu}$. One way 
is to proceed as follows: insert eq. (\ref{vbg_renorm}) into
(\ref{ren-cond}), and solve for $\bar{\nu}$, to get 
\begin{equation}
\bar{\nu}=\bar{g}^2(1- {m_{bf} \Lambda V_{bg} \over \pi^2})( {m_{bf}
  \Lambda \over 
  \pi^2}+ {\nu \over g}). 
\end{equation} 

From inspecting the above equation we can choose a
definition of $\bar{g}$, which will also imply one for $\bar{\nu}$,
and we get (reporting also eq.(\ref{vbg_renorm}) for completeness)
%\begin{widetext}
\begin{eqnarray} 
&&\bar{V_{bg}}=V_{bg} \left(1 \over 1- {m_{bf} \Lambda V_{bg} \over \pi^2}
\right)\nonumber \\
&&\bar{g}=g \left(1 \over 1- {m_{bf} \Lambda V_{bg} \over \pi^2}
\right)\nonumber \\ 
&&\bar{\nu}=\nu+\bar{g} g {m_{bf} \Lambda V_{bg} \over \pi^2}
\end{eqnarray}
%\end{widetext}
Using these definitions of $\bar{V_{bg}}$, $\bar{g}$ and $\bar{\nu}$,
together with the 
policy of 
imposing $\Lambda$ as the upper limit of momentum integrals, will
guarantee that observables will not depend on the choice of
$\Lambda$, as long as it is chosen to be bigger than momentum scales
relevant to experiment.
\subsubsection{Many-Body Generalization}

Generalizing the above treatment from two to many particles, we must
now account for the fact that, in a many-body system the molecular
self energy is modified by the environment. Unlike in the scattering
problem, the procedure outlined in the previous section is 
only an approximation to the full many body problem, but as some
evidence seems to suggest, a pretty good one \cite{pieri,montecarlo}.

To perform this generalization, one needs to calculate the
many body self energy using the many body free green functions
$D^0_{MB}(\omega,p)$ and $G^0_{MB}$, respectively  for bosons and
fermions, defined as \cite{negele}
\begin{widetext}
\begin{eqnarray}
&&D^0_{MB}(\omega,p)=|\phi|^2+{1 \over \hbar \omega - \epsilon_p^B +
  \mu_B + i0^+} \\
&&G^0_{MB}(\omega,p)={1 \over \hbar \omega - \epsilon_p^F +
  \mu_B + i0^+sign(\epsilon_b^F -
  \mu_B)}, 
\end{eqnarray}
where $\mu_{B(F)}$ is the bosonic (fermionic) chemical
potential. The molecular self energy becomes
\begin{eqnarray}
&&\Sigma^MB(\Omega,{\bf p})=-i g^2 \int {d\omega \over 2\pi} {d^3q \over (2\pi)^3}
D_0({\bf q},\omega)G_0({\bf p-q},\Omega-\omega) = \nonumber \\
&&\ \ \ \Sigma^{MB}_0(\Omega,p)+\Sigma^{MB}_\delta(\Omega,p)= 
-i g^2 |\phi|^2 G_0(\Omega,{\bf p}) 
  \nonumber \\&&  
\ \ \ + g^2\int_{p-kf}^{\Lambda} {d^3q \over
  (2\pi)^3} {1 \over \Omega -\epsilon^B_{\bf q} -\epsilon^F_{\bf p-q}
  +\mu_F + \mu_B};
\label{selfmb}
\end{eqnarray}
\end{widetext}
the two terms in this expression represent contributions from
condensed ($\Sigma^{MB}_0(\Omega,p)$) and noncondensed
($\Sigma^{MB}_\delta(\Omega,p)$)bosons, respectively. 

The (approximate) many body self energy in eq. \ref{selfmb} can be
easily shown to reduce to its correct two body counterpart defined in
eq. (\ref{M-E}), when the densities and chemical potentials are set to zero.
 However, if the contribution due to the non-condensed bosons
 $\Sigma^{MB}_{\delta}$ is omitted, then $\Sigma^{MB}$ clearly
 vanishes in the two-body limit, $\phi\rightarrow 0$.
There would then be no renormalization of the molecular propagator,
and the pole of the T-matrix would coincide with the bare
detuning, as shown in fig. \ref{freq}.

We remind the reader that neglecting the non-condensed component
of the bosonic field was a perfectly well justified approximation of
eqs.(\ref{eqm1})-(\ref{eqm9}), which implies that those equations are
already inadequate to reproduce the two body binding energies in the
low density limit. Indeed the part of resonant term in the
Hamiltonian 
containing the bosonic fluctuations vanishes
according to Wicks theorem, since it is an average of a three operator
correlation with respect to a density matrix which is Gaussian, in the
HFB approximation. To correct this problem one should extend the HFB
approximation and explicitly include three, and possibly higher,
particle cumulants, finding 
some other way to truncate the BBGKY hierarchy. The subtleties
involved in such a calculation, however, are many, and non trivial,
and will be the subject of further work.

\section{Conclusions} 
We have performed a study the non-equilibrium behavior in Bose-Fermi
mixtures subject to an interspecies Feshbach resonance, Using the HFB
approximation. We have found that this approximation is not adequate
to describe the system, which is quite remarkable since it has become
one of the standard approaches to resonant cold atom physics due to
its successes in Bose gases and two component Fermi gases.

The reason of this failure is found in the way in which the
theory treats non-condensed bosons.
This problem could be corrected by the explicit
inclusion of three (and possibly higher) point cumulants, which will
allow for a mechanism 
through which bosons and fermions could correlate to form
molecules. This task, however is beyond the scope of the current investigation.

\acknowledgements
We gratefully acknowledge the support of the DOE and the NSF, as
well as the W. M. Keck Foundation.  We acknowledge useful discussions
with J. N. Milstein, J. Wachter and M. J. Holland.

\appendix
\section{}
\label{Appendix1}
In this appendix we will present a sample derivation of one of the
equations of motion, namely that for $<\delta^{\dagger}\delta>$.

Starting with the Hamiltonian in coordinate space 
\begin{eqnarray}
H= &&\int dx \ \psi^{\dagger}(x) T^F(x) \psi(x)+ \nonumber \\
&&\int dx \ \phi^{\dagger}(x) T^B(x) \phi(x)+\nonumber \\
&&\int dx \ \xi^{\dagger}(x) T^M(x) \xi(x)+\nonumber \\
&&{1 \over 2}\ \gamma \int dx \  |\phi(x)|^4+\nonumber \\
&&V_{bg} \int dx \ |\phi(x)|^2 |\psi(x)|^2+ \nonumber \\
&&g \int dx\ (\xi^{\dagger}(x) \phi(x) \phi(x)+c.c) \nonumber \\
\end{eqnarray},
where $T^{\alpha}(x)$ is the kinetic energy of molecules, bosons or
fermions.

We then write the bosonic field in terms of its average and
fluctuations around it $\phi(x) = \phi_0(x)+\delta(x)$, where $\phi_0$
is a complex number. Inserting this
expression in the Hamiltonian, we get the following 
\begin{eqnarray}
H&&=E_0+\int dx \ \psi^{\dagger}(x) (T^F(x)+V_{bg}|\phi_0(x)|^2)
\psi(x)+ \nonumber \\ 
&&\int dx \ \delta^{\dagger}(x) T^B(x) \delta(x)+
\int dx \ \xi^{\dagger}(x) T^M(x) \xi(x)+\nonumber \\
&&{ \gamma}\int dx \ (4 |\phi_0(x)|^2 |\delta
(x)|^2+{\phi_0^*(x)}^2\delta(x)\delta(x)+\nonumber \\
&&\ \ \ \ \ \ \ \ \ \phi_0(x)^2\delta^{\dagger}(x)\delta^{\dagger}(x)) 
+\nonumber \\ 
&&\int dx \
\left(\phi_0^*(x)\delta(x)+\phi_0(x)\delta^{\dagger}(x)\right) \times \nonumber \\
&&\ \ \ \ \ \ \ \ \ \left({\gamma\over 2} |\phi_0(x)|^2+V_{bg}|\psi(x)|^2\right)+ \nonumber \\
&&{\gamma} \int dx \
\left(\phi_0^*(x)\delta^{\dagger}(x)\delta(x)\delta(x) + c.c \right)+
\nonumber\\ 
&&{\gamma \over 2}\int dx \
\delta^{\dagger}(x)\delta^{\dagger}(x)\delta(x)\delta(x) +\nonumber\\
&&V_{bg} \int dx \ |\delta(x)|^2 |\psi(x)|^2+ \nonumber \\
&&g \int dx \ \left[\xi(x)^{\dagger} \left(\phi_0(x)+\delta(x)\right)
  \phi(x)+c.c \right],\nonumber \\
\end{eqnarray}
where $E_0$ is a constant which depends on $ \phi_0$, and its relevant for its
motion, but does not contribute to that of $\delta^{\dagger}\delta$.

The next step is to calculate the commutator
$[\delta^{\dagger}(z)\delta(z'),H]$, and to take its average, thereby
obtaining 

\begin{widetext}
\begin{eqnarray}
<[\delta^{\dagger}(z)\delta(z'),H]>=&&\left(T^B(z')-T^B(z)\right)<\delta^{\dagger}(z)\delta(z')>+
\nonumber \\
&&\gamma \left[2|\phi_0(z')|^2<\delta^{\dagger}(z)\delta(z')>+\phi_0^2(z)
  <\delta^{\dagger}(z)\delta^{\dagger}(z')>-\right.\nonumber \\
&&\left. 2|\phi_0(z)|^2<\delta^{\dagger}(z)\delta(z')>-{\phi_0^*}^2(z')
  <\delta(z)\delta(z')> \right]+\nonumber \\
&&\phi_0(z')\left( {\gamma}<\delta^{\dagger}(z')>|\phi_0(z')|^2+V_{bg}
\int dx \ <\delta^{\dagger}(z')\psi^{\dagger}(x)\psi(x)>\right)- \nonumber \\
&&\phi_0^*(z)\left({\gamma}<\delta(z)>|\phi_0(z)|^2+V_{bg} \int dx \
<\delta(z)\psi^{\dagger}(z)\psi(z)>\right)+ \nonumber \\
&&{\gamma} \int dx \ \left[\phi_0^*(x)
  (<\delta^{\dagger}(z)\delta(z')\delta^{\dagger}(x)\delta(x)\delta(x)> 
-<\delta^{\dagger}(x)\delta(x)\delta(x)\delta^{\dagger}(z)\delta(z')>)+
\right. \nonumber \\ 
&&\left. \phi_0(x)
  (<\delta^{\dagger}(z)\delta(z')\delta^{\dagger}(x)\delta^{\dagger}(x)\delta(x)>  
-<\delta^{\dagger}(x)\delta^{\dagger}(x)\delta(x)\delta^{\dagger}(z)\delta(z')>)\right]
+\nonumber \\  
&&\gamma
\left(<\delta^{\dagger}(z)\delta^{\dagger}(z')\delta(z')\delta(z')>-
<\delta^{\dagger}(z)\delta^{\dagger}(z)\delta(z)\delta(z')>\right)+ 
\nonumber \\  
&&V_{bg}\left(
<\delta^{\dagger}(z)\delta(z')\psi^{\dagger}(z')\psi^(z')>-
<\delta^{\dagger}(z)\delta(z')\psi^{\dagger}(z)\psi^(z)> \right) + \nonumber
\\
&&g \left(<\xi(z')\psi^{\dagger}(z')\delta^{\dagger}(z)>-
<\xi^{\dagger}(z)\psi(z)\delta(z')> \right).
\end{eqnarray}
\end{widetext}

The next step is to apply Wick's theorem to correlation functions of
three or more operators. This implies that all correlation functions
of odd order will vanish.We then get 
\begin{widetext}
\begin{eqnarray}
<[\delta^{\dagger}(z)\delta(z'),H]>=&&\left(T^B(z')-T^B(z)\right)<\delta^{\dagger}(z)\delta(z')>+
\nonumber \\
&&\gamma
\left[2|\phi_0(z')|^2<\delta^{\dagger}(z)\delta(z')>+ \right.\nonumber \\ 
&&\left. \phi_0^2(z)
  <\delta^{\dagger}(z)\delta^{\dagger}(z')>-2|\phi_0(z)|^2<\delta^{\dagger}(z)\delta(z')>-{\phi_0^*}^2(z')
  <\delta(z)\delta(z')> \right]+\nonumber \\
&&\gamma
\left(<\delta^{\dagger}(z)\delta^{\dagger}(z')><\delta(z)\delta(z')> +  
2<\delta^{\dagger}(z')\delta(z')><\delta^{\dagger}(z)\delta(z')> - \right.\nonumber \\ 
&&\left. 2<\delta(z)\delta^{\dagger}(z)><\delta^{\dagger}(z)\delta(z')> 
- <\delta^{\dagger}(z)\delta^{\dagger}(z)><\delta(z)\delta(z')>\right)
\nonumber \\  
&&V_{bg}\left(
<\delta^{\dagger}(z)\delta(z')><\phi^{\dagger}(z')\phi^(z')>-
<\delta^{\dagger}(z)\delta(z')><\phi^{\dagger}(z)\phi^(z)> \right) + \nonumber
\\
\end{eqnarray}
\end{widetext}

 In free space, $\phi_0$ becomes a constant, and all two point
 correlations, which are functions of z,z', become functions of z-z',
 so that in momentum space they become functions of a single
 momentum. We thus obtain eq. (\ref{eqm2}).

\end{document}